# Capacity-Approaching Signal Constellations for the Additive Exponential Noise Channel

Stéphane Y. Le Goff

School of Electrical, Electronic and Computer Engineering
Newcastle University, United Kingdom

***Abstract*** – We present a new family of signal constellations, called *log constellations*, that can be used to design near-capacity coded modulation schemes over additive exponential noise (AEN) channels. Log constellations are designed by *geometrically* approximating the input distribution that maximizes the AEN channel capacity. The mutual information achievable over AEN channels with both coded modulation (CM) and bit-interleaved coded modulation (BICM) approaches is evaluated for various signal sets. In the case of CM, the proposed log constellations outperform, sometimes by over half a decibel, the best existing signal sets available from the literature, and can display error performance within only 0.12 dB of the AEN channel capacity. In the context of BICM, log constellations do not offer significant performance advantages over the best existing constellations. As the potential performance degradation resulting from the use of BICM instead of CM is larger than 1 dB, BICM may however not be a suitable design approach over AEN channels.

## I. INTRODUCTION

The additive exponential noise (AEN) channel constitutes the natural discrete-time model for the continuous-time Gaussian channel when the transmitted signal is affected by a fast-varying phase noise [1]. The resulting lack of coherence between signal and noise components at the receiver input makes it impossible to detect the complex-valued quadrature amplitudes. In such case, information can only be transmitted by modulating the signal energy and using a direct (non-coherent) receiver architecture to detect it. As an example, the AEN channel can be considered as a good model for digital communication systems using electromagnetic radiation as a photon gas with no quantum interference effects [2].

In 1996, Verdú investigated various information-theoretical aspects of the exponential distribution and showed that the AEN channel shares a number of similarities with its additive white Gaussian noise (AWGN) counterpart [3]. In particular, he determined the expression of the AEN channel capacity and found it to be identical to that of an equivalent AWGN channel. More recently, Martinez considered the issue of designing coded modulation (CM) schemes for the AEN channel [1], [4]. One of his most interesting results was the introduction of a family of signal constellations that can perform approximately 0.76 dB away from capacity at high signal-to-noise ratio (SNR).

In this paper, we propose another family of signal sets, hereafter called *log constellations*, that can significantly outperform those introduced by Martinez when used for CM design. In particular, it is shown that, when the number of constellation symbols is greater than or equal to 1024, an error performance within only 0.12 dB of the capacity is achievable with a CM system using log constellations. We also consider the design of bit-interleaved coded modulation (BICM) systems based on the proposed log constellations. Recall that BICM is a low-complexity, sub-optimal alternative to coded modulation introduced by Zehavi in 1992 [5]. Due to its simplicity and near-optimal error performance over AWGN and fading channels, this approach has actually become the *de facto* standard for wireless communication systems (see, e.g., [6] and [7]). It is shown that, over AEN channels, the error performance is significantly degraded when employing BICM rather than CM. The obtained results also indicate that log constellations are not particularly attractive for BICM.

The remainder of the paper is organized as follows. In Section II, we provide a brief review of the AEN channel and describe the technique used to obtain the proposed log constellations. The mutual information achievable with the CM approach using log constellations is then evaluated in Section III for different numbers of constellation symbols and compared to the results obtained with Martinez's signal sets. In Section IV, similar evaluations and comparisons of mutual information are performed in the context of BICM. Finally, conclusions are drawn in Section V.

## II. A NEW FAMILY OF SIGNAL CONSTELLATIONS FOR AEN CHANNELS

### *A. AEN Channel Model*

Consider the transmission of a symbol $x \in \{x_1, \ldots, x_M\}$ drawn from a one-dimensional $M$-ary signal constellation composed of non-negative real symbols $x_1$, $x_2\ldots$, and $x_M$. At the output of an AEN channel, the corresponding sample $y$ is given by

$$y = x + n, \tag{1}$$

where $n$ designates a non-negative real noise sample with a probability density function (PDF)

$$P_n(x) = \frac{1}{E_n} \exp\left(-\frac{x}{E_n}\right) u(x). \tag{2}$$

In this expression, $E_n$ denotes the average value of the noise samples and $u(x)$ is the unit step function defined as follows: $u(x) = 1$ for $x \geq 0$, and $u(x) = 0$ for $x < 0$. The channel signal-to-noise ratio (SNR) $\gamma$ is defined as

$$\gamma = \frac{E_s}{E_n}, \tag{3}$$

where $E_s$ is the average value of the transmitted $M$-ary symbols $x$. To simplify the notations used throughout this paper, we will assume from now on that the signal constellation has unit average value, i.e. $E_s = 1$ and thus $\gamma = (E_n)^{-1}$.

In his 1996 paper, Verdú showed that the capacity of the AEN channel (expressed in bits/channel use) is given by

$$C = \log_2(1+\gamma), \tag{4}$$

and is therefore identical to that of an equivalent AWGN channel [3]. In order to achieve this capacity, the channel input $x$ must have the following distribution:

$$P_x(x) = \frac{E_s}{(E_s + E_n)^2} \exp\left(-\frac{x}{E_s + E_n}\right) + \frac{E_n}{E_s + E_n} \delta(x), \qquad x \geq 0, \tag{5}$$

which is, with the notations used in this paper, equivalent to

$$P_x(x) = \left(\frac{\gamma}{\gamma+1}\right)^2 \exp\left(-\frac{\gamma}{\gamma+1}\,x\right) + \frac{1}{\gamma+1}\,\delta(x), \qquad x \ge 0, \tag{6}$$

where $\delta(x)$ denotes the Dirac delta function.

*B. Signal constellations for the AEN Channel*

Eqn. (6) clearly implies that, in AEN channels, signal constellations that are composed of equiprobable uniformly spaced symbols cannot achieve the capacity given by (4). Actually, it was shown in [1] and [4] that, at high SNR, the gap between channel capacity and uniformly spaced equiprobable signaling is approximately equal to 1.33 dB. It is worth mentioning that this SNR gap is slightly smaller than that obtained at high SNR over AWGN channels, the latter being approximately equal to 1.53 dB [8].

Constellation shaping must be used in order to reduce the SNR gap. Perhaps the simplest method to implement constellation shaping is called *geometrical shaping* and consists of using an equiprobable signal set for which the symbols are not uniformly spaced on the one-dimensional axis. In particular, it has been shown that *geometrical Gaussian-like M*-ary signal constellations can achieve the channel capacity of the AWGN channel as $M \to +\infty$ [9]. In [1] and [4], Martinez proposed a method to implement geometrical shaping for the AEN channel. Basically, he introduced a family of $M$-ary signal sets composed of equiprobable non-negative real symbols $x_i$, $i \in \{1, 2,\ldots, M\}$, defined as follows:

$$x_i = \beta\,(i-1)^{\lambda}, \tag{7}$$

where $\lambda$ is a design parameter that can be set to any value and $\beta$ is simply a normalization parameter computed so that the average value of the symbols $x_i$ is equal to the unit, i.e. $E_s = 1$. Martinez was able to show that operation at high SNR at approximately 0.76 dB away from the AEN channel capacity can be achieved when $\lambda = \left(1+\sqrt{5}\right)/2 \approx 1.618$, which corresponds to a shaping gain of approximately 0.57 dB over uniformly spaced equiprobable signaling.

In this paper, we propose a new family of signal sets that outperform the constellations proposed in [1] and [4] over the AEN channel. In order to design these sets, the idea is to *geometrically* implement the input distribution given in (6), i.e. place the real symbols $x_i$, $i \in \{1, 2, \ldots, M\}$, over the interval $[0,+\infty]$ according to (6). Unfortunately, it appears that the discrete component represented by the Dirac delta function in (6) actually makes this goal impossible to achieve in practice. This is the reason why we have to replace (6) with the following continuous PDF:

$$P_x(x) = \frac{\gamma}{2\,(\gamma+1)} \exp\left(-\frac{\gamma}{\gamma+1}|x|\right), \tag{8}$$

which is very similar to (6). Note that $P_x(x)$ is now an even function which takes non-zero values over the entire interval $[-\infty,+\infty]$.

The initial step consists of dividing the one-dimensional axis into $(2M-1)$ intervals $[\alpha_k, \alpha_{k+1}]$ chosen so that the areas under the function $P_x(x)$ are equal for all these intervals:

$$\int_{\alpha_k}^{\alpha_{k+1}} P_x(x)\,dx = \frac{1}{2M-1}, \qquad k \in \{1, \ldots, 2M-1\}. \tag{9}$$

By combining (8) and (9), and using $\alpha_1 = -\infty$ and $\alpha_{2M} = +\infty$, we show that the parameters $\alpha_k$ can be expressed as

$$\alpha_k = \frac{\gamma+1}{\gamma} \ln\left(\frac{2\,(k-1)}{2M-1}\right), \qquad \text{for } k \in \{1, \ldots, M\}, \tag{10}$$

and

$$\alpha_k = -\frac{\gamma+1}{\gamma} \ln\left(\frac{2\,(2M-k)}{2M-1}\right), \qquad \text{for } k \in \{M+1, \ldots, 2M-1\}, \tag{11}$$

where $\ln(\cdot)$ designates the natural logarithm.

Then, we define $(2M-1)$ signal points $x_k$ as the centroids of the intervals $[\alpha_k, \alpha_{k+1}]$ with respect to the distribution $P_x(x)$ given in (8):

$$x_k = (2M-1)\int_{\alpha_k}^{\alpha_{k+1}} x\, P_x(x)\, dx, \qquad k \in \{1, ..., 2M-1\}. \tag{12}$$

For $k \in \{M+1, ..., 2M-1\}$, the evaluation of (12) yields

$$x_k = \frac{\gamma+1}{\gamma}\left[f(k) - f(k+1)\right], \tag{13}$$

where $f(\cdot)$ is a function defined as

$$f(x) = (2M-x)\ln\left(\frac{e(2M-1)}{2(2M-x)}\right). \tag{14}$$

Finally, an *M*-ary signal set suitable for the AEN channel is generated by simply keeping the *M* symbols $x_k$, $k \in \{M, ..., 2M-1\}$, and discarding the symbols $x_k$, $k \in \{1, ..., M-1\}$. We can easily show that $x_M = 0$ whereas the values of the $(M-1)$ other symbols $x_k$, $k \in \{M+1, ..., 2M-1\}$ are computed using (13) and (14).

To simplify the notations, we can at this stage replace the index *k* with an index *i* defined as $i = k - M + 1$. The set of *M* symbols $x_k$, $k \in \{M, ..., 2M-1\}$, thus becomes a set of *M* symbols $x_i$, $i \in \{1, ..., M\}$, defined as follows:

$$x_1 = 0 \text{ and } x_i = \beta\left[g(i) - g(i+1)\right],\ i \in \{2, ..., M\}, \tag{15}$$

where $\beta$ is a normalization parameter computed so that the average value of the symbols $x_i$ is equal to the unit, i.e. $E_s = 1$, and $g(\cdot)$ is a function defined as

$$g(x) = (M+1-x)\ln\left(\frac{e(2M-1)}{2(M+1-x)}\right). \tag{16}$$

As $M \to +\infty$, the constellation defined by (15) and (16) actually implements a geometrical version of (8) over the interval $[0,+\infty]$ and can thus be used over AEN channels to design capacity-approaching CM systems. Note that, hereafter, constellations obtained using (15) and (16) will be referred to as *log constellations* as the signal points are defined using the logarithm function.

## III. MUTUAL INFORMATION ACHIEVABLE WITH CM SYSTEMS USING LOG CONSTELLATIONS

In this Section, we consider the design of coded modulation (CM) systems using the log constellations introduced in Section II. Let $x$ and $y$ denote respectively the transmitted symbol and the corresponding received sample after transmission over the AEN channel. Information theory tells us that the highest rate, expressed in bits/channel use, at which information can be transmitted reliably using a CM scheme is given by the mutual information

$$I = E_{x,y}\left[\log_2\left(\frac{p(y|x)}{\sum_{i=1}^{M} p(y|x_i)\, p(x_i)}\right)\right], \tag{17}$$

where the operator $E_{x,y}[\,\cdot\,]$ designates the expectation of "$\cdot$" with respect to $x$ and $y$, $p(x_i)$ is the probability of transmitting a particular symbol $x_i$, $i \in \{1, ..., M\}$, whereas $p(y|x)$ and $p(y|x_i)$ are the probability density functions of the sample at the channel output given the transmission of symbols $x$ and $x_i$, respectively. It can easily be shown that, for equiprobable signaling over the AEN channel, (17) can be written as

$$I = E_{x,y}\left[\log_2\left(\frac{M\,\exp(-\gamma\,(y-x))}{\sum_{i=1}^{M}\exp(-\gamma\,(y-x_i))\,u(y-x_i)}\right)\right]. \tag{18}$$

We have evaluated (18) for several $M$-ary log constellations, with $M$ ranging from 4 to 256, using numerical integration via the Monte Carlo method. Fig. 1 shows the variation of the mutual information $I$ as a function of the SNR $\gamma = E_s / E_n$ for some of these constellations. For comparison sake, we have also plotted in Fig. 1 the AEN channel capacity computed using (4) as well as the results obtained with the constellations proposed by Martinez and defined using (7) with $\lambda$ = 1.618 [1], [4]. Note that, to be able to display results clearly using a reasonable amount of space, we decided not to show in Fig. 1.a the plots obtained with $M$ = 8, 32, and 128.

It is seen from Fig. 1 that, for all mutual information values, the log constellations achieve a better performance than those proposed by Martinez in the sense that they are able to perform closer to the AEN channel capacity. For instance, for $I$ = 1 bit/channel use, we observe that the 16-ary log constellation is able to outperform the best Martinez's signal set by approximately 0.29 dB. If the desired mutual information is greater than 1.5 bit/channel use, the best log signal set is always the one with $M$ = 256 symbols. For $I$ = 2 and 3 bits/channel use, the SNR gaps between this constellation and the best Martinez's signal set are about equal to 0.34 dB and 0.44 dB, respectively. When mutual information values of 4 and 5 bits/channel use are desired, the 256-ary log constellation outperforms the best Martinez's signal set by approximately 0.40 dB and 0.37 dB, respectively. It is worthwhile mentioning that, for $I$ = 3 and 4 bits/channel use, Fig. 1 indicates that a CM scheme designed using a 256-ary log constellation is actually capable of operating within only 0.23 dB of the capacity limit.

We remark that increasing the number $M$ of constellation symbols clearly has a beneficial effect on the performance achieved by log constellations, which is not the case with Martinez's signal sets. This can merely be explained as follows: As $M \to +\infty$, the geometrical implementation of (8) becomes more and more accurate, thus progressively bringing the mutual information achievable with a log constellation closer to the capacity limit. In order to further clarify this point, we have plotted in Fig. 2 the variation of the mutual information $I$ as a function of the SNR $\gamma = E_s / E_n$, for a number $M$ of constellation symbols ranging from 256 to 2048 and a desired value of $I$ around 4 bits/channel use.

Fig. 2 clearly indicates that increasing $M$ from 256 to 2048 does not result in any performance improvement when the constellations proposed by Martinez are employed. As an example, for $I = 4$ bits/channel use, we can see that an error performance within about 0.64 dB of the capacity is achievable when using any of the four Martinez's constellations considered here. If the log signal sets are employed, we achieve a SNR gain of about 0.1 dB at $I = 4$ bits/channel use as $M$ is increased from 256 to 1024. However, it appears that no further improvement can be achieved by increasing the number of constellation symbols beyond 1024. In any case, these results show that a CM scheme designed using log constellations with $M \geq 1024$ is able to perform approximately 0.12 dB away from the capacity limit of the AEN channel. This corresponds to a significant SNR gain of about 0.52 dB over an equivalent CM system employing a Martinez's signal set.

## IV. MUTUAL INFORMATION ACHIEVABLE WITH BICM SYSTEMS USING LOG CONSTELLATIONS

We now consider the design of bit-interleaved coded modulation (BICM) systems using the log constellations introduced in Section II. BICM is a sub-optimal alternative to CM that has become very popular over the last decade or so owing to its low complexity and near-optimal error performance over AWGN and fading channels [5] – [7]. The idea behind BICM is to map the encoded bits, after interleaving, to a certain constellation using Gray mapping. The decoding is performed by first computing the log-likelihood ratios of the coded bits, and then, after de-interleaving, using a binary decoder as if these log-likelihood ratios were the observations at a binary phase-shift keying/quaternary phase-shift keying (BPSK/QPSK) channel output.

Due to the importance of BICM for practical applications, it is interesting to investigate the potential of log constellations for BICM design over AEN channels. Consider an *M*-ary modulation modelled by a signal set S composed of $M$ symbols. Let $c = (c_1, c_2, ..., c_m) \in \{0, 1\}^m$ denote a vector of $m = \log_2(M)$ coded bits at the modulator input, and $y$ the corresponding received sample at the channel output. It was shown in [6] that the generic expression of the BICM mutual information *I*, valid under the constraint of uniform input distribution, is given by

$$I = \sum_{i=1}^{m} E_{c,y}\left[1 + \log_2\left(\frac{\sum_{x \in \mathrm{S}_{i,c_i}} p(y|x)}{\sum_{x \in \mathrm{S}} p(y|x)}\right)\right], \tag{19}$$

where $E_{c,y}[\,\cdot\,]$ designates the expectation of "$\cdot$" with respect to $c$ and $y$, and $\mathrm{S}_{i,c_i}$ denotes the subset of all constellation symbols $x \in \mathrm{S}$ whose labels have the value $c_i \in \{0,1\}$ in position *i*. Assuming transmission over an AEN channel, we can easily show that (19) is equivalent to

$$I = m + E_{c,y}\left[\log_2\left(\frac{\prod_{i=1}^{m}\sum_{x\in S_{i,c_i}}\exp\left(-\gamma\left(y-x\right)\right)u\left(y-x\right)}{\left(\sum_{x\in S}\exp\left(-\gamma\left(y-x\right)\right)u\left(y-x\right)\right)^m}\right)\right]. \tag{20}$$

By performing a numerical integration via the Monte Carlo method, we have evaluated (20) in order to determine the highest rate at which information can be transmitted reliably with the BICM approach. We have considered the use of various $M$-ary log constellations, with $M$ ranging from 4 to 256. Fig. 3 shows the variation of the BICM mutual information $I$ as a function of the SNR $\gamma = E_s / E_n$ for some of these constellations. As we did for the CM case, we have also plotted in Fig. 3 the curves obtained with Martinez's constellations, defined using (7) with $\lambda$ = 1.618, as well as the AEN channel capacity given by (4). Once again, to be able to display results clearly using a reasonable amount of space, we have chosen to display in Fig. 3 only the most interesting results among all those which have been obtained.

It is seen from Fig. 3.a that, for low mutual information values, the best performance is achieved using Martinez's constellations with a small number of symbols. For example, for $I$ = 1 and 2 bits/channel use, the best signal sets are those defined using (7) with $M$ = 4 and 8 symbols, respectively, and they outperform the best log constellations by approximately 0.15 dB and 0.20 dB, respectively. However, as the desired mutual information value is increased beyond a value of 3 bits/channel use, the log constellations tend to perform marginally better than Martinez's signal sets. As an example, for $I$ = 4, 5, and 6 bits/channel use, we observe that the best constellation is always the 256-ary log signal set, the latter being, in all three cases, able to outperform the best Martinez's signal set by approximately 0.10-0.15 dB.

These results indicate that, in the particular context of BICM design, the log constellations introduced in this paper do not offer any significant performance advantage over the signal sets proposed by Martinez. It is worthwhile noting that, for all constellations considered here, the SNR gap between the AEN channel capacity and the achievable mutual information is always greater than about 1.40 dB, which is much larger than the typical gap values obtained with the CM approach. We can thus conclude that, over AEN channels, BICM may not really be considered as a near-optimal technique to combine modulation and coding.

## V. CONCLUSIONS

We have introduced a new family of signal constellations, referred to as *log constellations*, that can be employed for designing near-capacity CM schemes over AEN channels. For these signal sets, the distribution of the symbols on the one-dimensional axis is a *geometrical* approximation of the optimal input distribution determined by Verdú in 1996 [3]. We have evaluated the mutual information achievable over AEN channels with both CM and BICM approaches for various signal sets. The results have shown that, in the case of CM, the proposed log constellations significantly outperform, by over half a decibel in some cases, the best existing signal sets available from the literature. We have also shown that operation within only 0.12 dB of the AEN channel capacity is possible when using log constellations with a number of symbols greater than or equal to 1024. In the context of BICM, it has been observed that log constellations do not offer significant performance advantages over the best existing constellations. In any case, we believe that, despite its simplicity, BICM may not be a suitable approach to modulation and coding over AEN channels. In fact, our results have indicated that the potential performance degradation resulting from the use of BICM instead of CM is larger than 1 dB, which implies that BICM may not really be considered as a *near-optimal* approach over AEN channels.

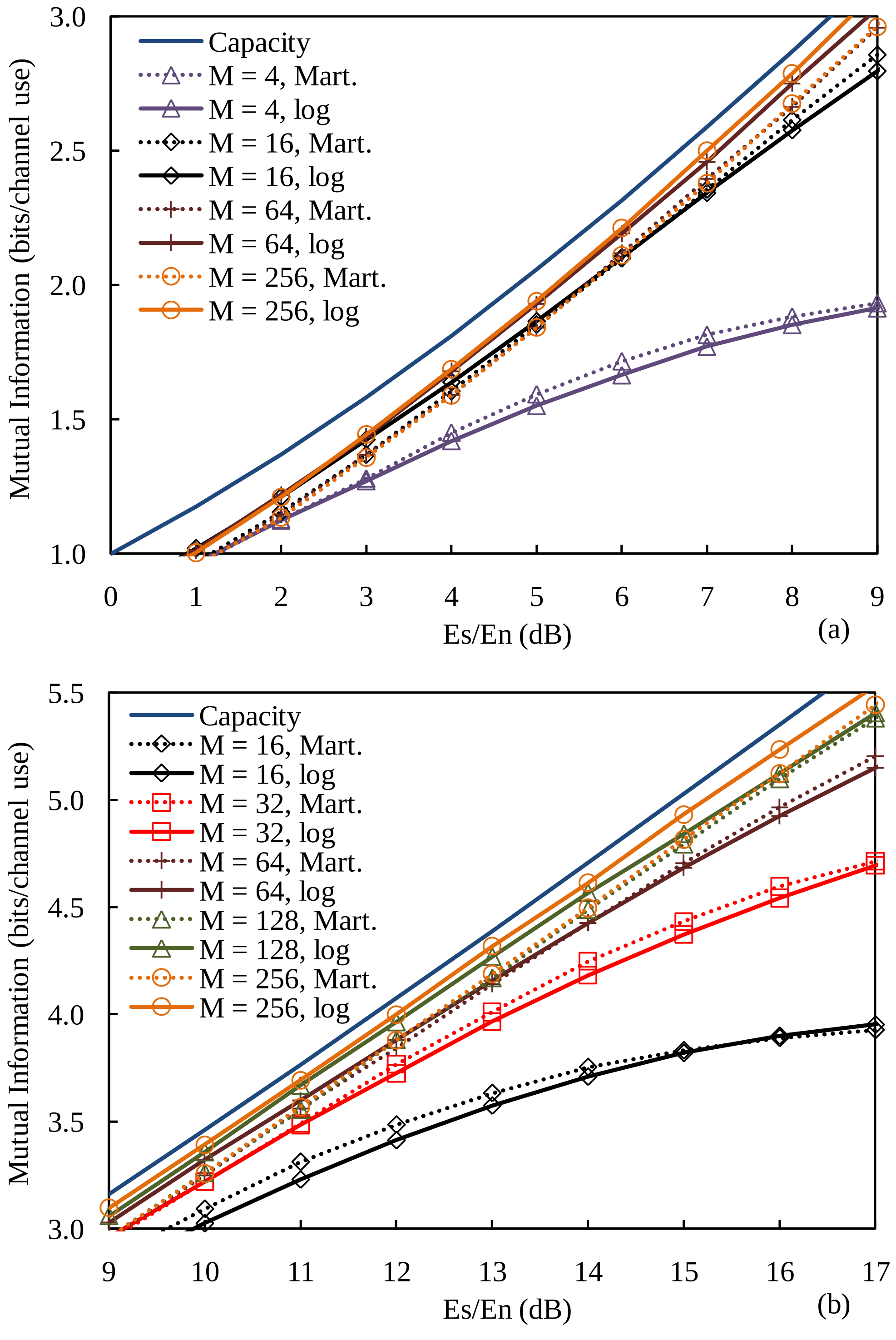

Fig. 1 – Mutual information achievable with a CM scheme as a function of the SNR, for various log constellations (“log”), over the AEN channel. For comparison purposes, we have also plotted the curve corresponding to the channel capacity as well as the mutual information curves obtained with the signal sets proposed by Martinez (“Mart.”). The number $M$ of constellation symbols ranges from 4 to 256. For clarity sake, we decided not to show the plots obtained for $M$ = 8, 32, and 128 in Fig. 1.a.

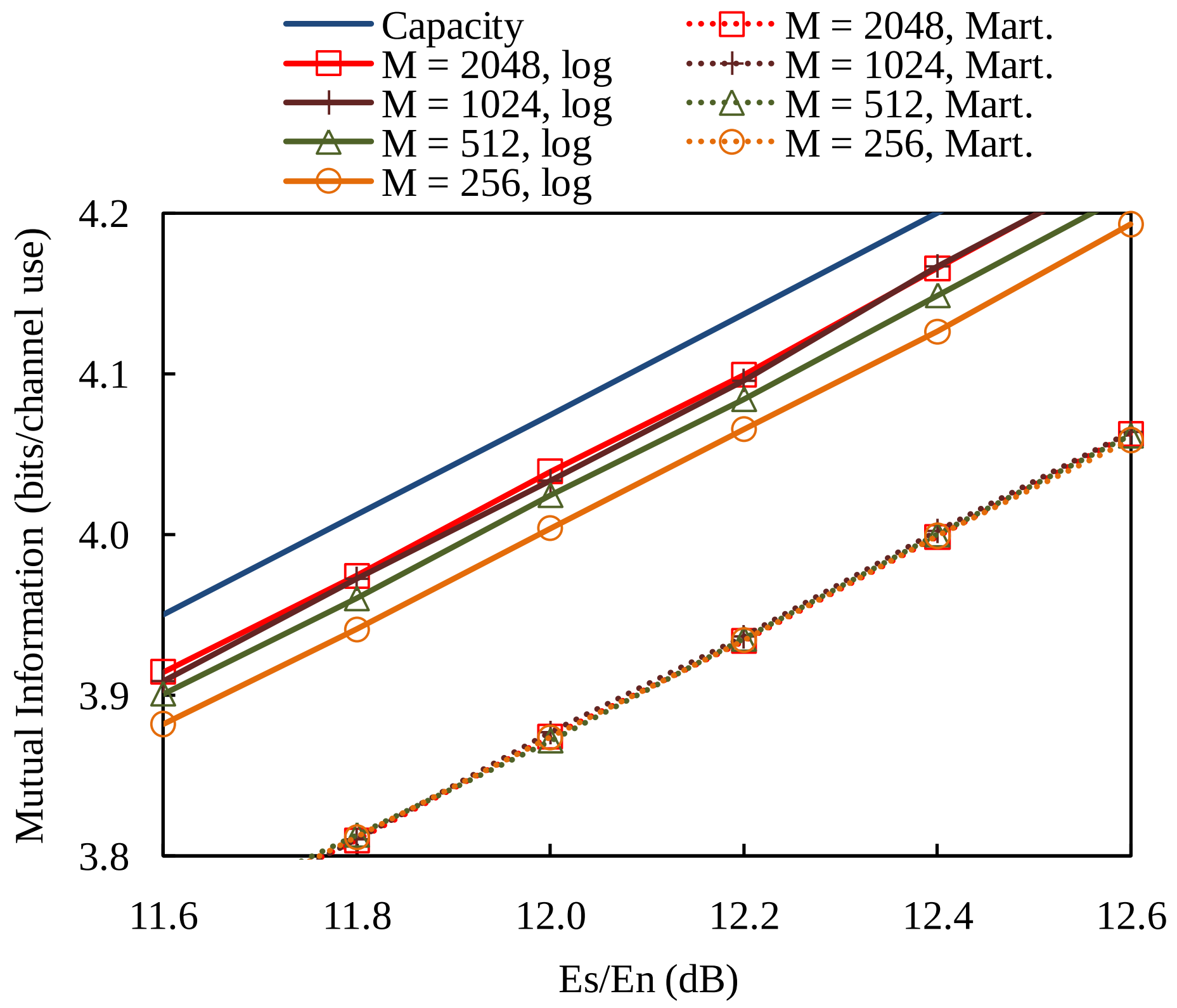

Fig. 2 – Mutual information achievable over the AEN channel with a CM scheme as a function of the SNR, for various log constellations ("log") and desired mutual information values around 4 bits/channel use. For comparison purposes, we have also plotted the curve corresponding to the AEN channel capacity as well as the mutual information curves obtained with the signal sets proposed by Martinez ("Mart."). The number *M* of constellation symbols ranges from 256 to 2048.

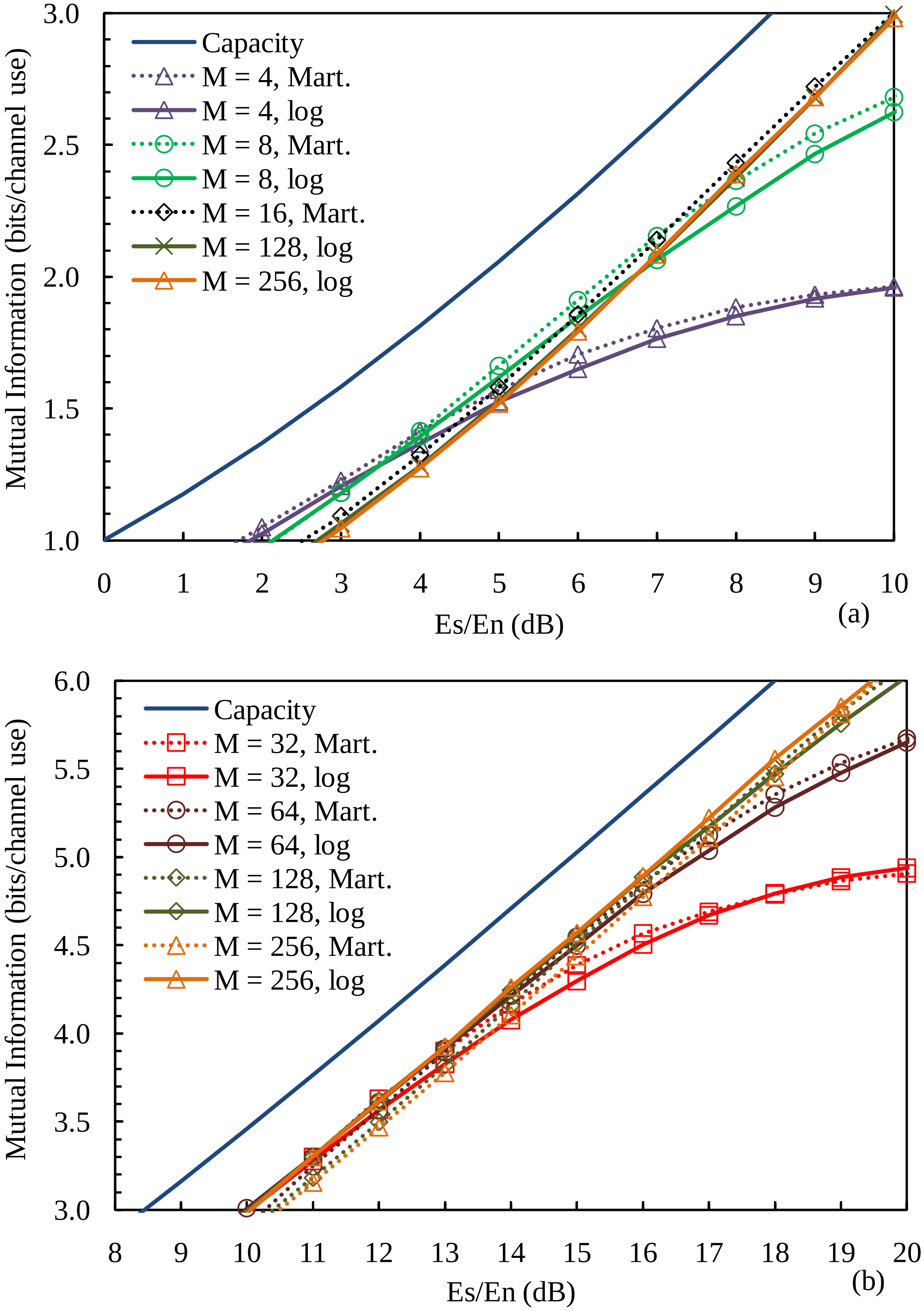

Fig. 3 – Mutual information achievable with a BICM scheme as a function of the SNR, for various log constellations ("log"), over the AEN channel. For comparison purposes, we have also plotted the curve corresponding to the channel capacity as well as the mutual information curves obtained with the signal sets proposed by Martinez ("Mart."). The number *M* of constellation symbols ranges from 4 to 256. For clarity sake, we decided not to show some curves that were of no particular interest.